# Effects of paramagnetic pair-breaking and spin-orbital coupling on multi-band superconductivity

Yilikal Ayino[1], Jin Yue[2], Tianqi Wang[2], Bharat Jalan[2] and Vlad S. Pribiag[1*]

[1]School of Physics and Astronomy, University of Minnesota
[2]Department of Chemical Engineering and Materials Science, University of Minnesota
[*]Corresponding author: vpribiag@umn.edu

## Abstract

The BCS picture of superconductivity describes pairing between electrons originating from a single band. A generalization of this picture occurs in multi-band superconductors, where electrons from two or more bands contribute to superconductivity. The contributions of the different bands can result in an overall enhancement of the critical field and can lead to qualitative changes in the temperature dependence of the upper critical field when compared to the single-band case. While the role of orbital pair-breaking on the critical field of multi-band superconductors has been explored extensively, paramagnetic and spin-orbital scattering effects have received comparatively little attention. Here we investigate this problem using thin films of Nd-doped $SrTiO_3$. We furthermore propose a model for analyzing the temperature-dependence of the critical field in the presence of orbital, paramagnetic and spin-orbital effects, and find a very good agreement with our data. Interestingly, we also observe a dramatic enhancement in the out-of-plane critical field to values well in excess of the Chandrasekhar-Clogston (Pauli) paramagnetic limit, which can be understood as a consequence of multi-band effects in the presence of spin-orbital scattering.

## INRODUCTION

In the conventional picture of superconductivity electrons that form the Cooper pairs originate from a single band. However, in certain materials superconductivity involves electrons from multiple bands, leading to multiple, interacting superconducting condensates. The coherent interactions between the condensates in multi-band superconductors are predicted to give rise to unique effects, such as vortices with fractional flux[1], time-reversal breaking solitons[2], and enhanced upper critical fields ($H_{c2}$) [3]. Experimental evidence for multi-band superconductivity has been reported in several materials, including $MgB_2$,[4] iron-based superconductors[5,6], the heavy-fermion superconductor $PrOs_4Sb_{12}$[7], and $NbSe_2$[8]. The majority of these are bulk, three-dimensional superconductors. Here we report on the interplay between multi-band superconductivity and spin-orbital (SO) scattering in thin films of doped $SrTiO_3$. Superconductivity in STO (which in its undoped form is a band insulator) remains a rich and widely-studied problem over half a century since its initial discovery[9].

The band structure of STO is particularly interesting because up to three bands can contribute to conduction, depending on doping levels. Specifically, the conduction band of STO is composed of three $t_{2g}$ orbitals originating from titanium *d* bands. The degeneracies of the three $t_{2g}$ orbitals are lifted by SO coupling and by the crystal field, in the low temperature tetragonal phase[10]. As a result, a natural question is whether superconductivity in STO can also involve multiple bands. The earliest report of possible two-band superconductivity in STO comes from a double-peak feature observed in tunneling spectroscopy on Nb-doped STO above a certain carrier concentration[11], which could indicate two superconducting gaps. Other work has shown that the transition temperature ($T_c$) peaks near Lifshitz transitions between STO bands[12], which could indicate that multiple bands are involved in STO superconductivity. On the other hand, more recent tunneling spectroscopy studies on thin films of Nb-doped STO have reported only a single coherence peak as a function of tunneling bias for a wide range of carrier concentrations, consistent with single-band superconductivity[13]. In addition, microwave radiation experiments on bulk Nb-doped STO have been found to be consistent with single-gap superconductivity[14]. Similarly, in superconducting $LaAlO_3/SrTiO_3$ interfaces, weakening of superfluid stiffness near $T_c$ at large charge densities has been attributed to the onset of multi-band superconductivity[15], yet SQUID measurements of the superfluid density vs. temperature have also observed results consistent with only a single-band BCS picture[16]. One possible reason for these discrepancies is different degrees of interband-coupling in different samples or gap homogenization due to defect scattering[14], which can affect the visibility of the multi-band character. Here we investigate multi-band superconductivity in thin films of doped STO using the temperature-dependence of the upper critical field ($H_{c2}$). This approach is well-suited precisely for the dirty limit, where disorder may limit the sensitivity of other experimental probes[17].

Multi-band superconductivity can result in marked differences in the temperature dependence of $H_{c2}$ vs. the single-band case, as well as in an overall enhancement of $H_{c2}(T)$, as observed for example in Fe-based superconductors[5] and $MgB_2$[3,18,19]. However, previous theoretical work on multi-band superconductivity has so far not considered the combined effects on $H_{c2}(T)$ of orbital depairing, paramagnetic depairing and SO scattering[3,17,20]. We propose an extension of existing two-band superconductivity models to include all these effects. As described below, we find very good quantitative agreement of our extended model with our data on doped STO. This model is not specific to STO and we expect it to be applicable to other multi-band superconductors in which SO scattering effects are important.

## EXPERIMENTAL DETAILS

To date, STO has been shown to superconduct when doped with one of a handful of dopants: O vacancies[9], Nb[21], La[22] and Sm[23], with critical temperatures in the range of few hundred mK. In addition, superconductivity has been reported in ionic-liquid-gated STO[24], with a similar range of critical temperatures as chemically-doped doped STO. Our samples are grown using a previously-unexplored dopant, $Nd^{3+}$, which substitutes for $Sr^{2+}$ (forming $Nd_xSr_{1-x}TiO_3$) and therefore n-dopes the STO, as confirmed by Hall measurements. Interestingly, unlike all previous STO dopants, $Nd^{3+}$ has a net spin, suggesting that Nd could act as a magnetic dopant, although this is not the focus of this work.

The Nd-doped $SrTiO_3$ films in this study were grown homoepitaxially on a thin undoped buffer $SrTiO_3$ film on $SrTiO_3$ (001) substrate (Crystec GmbH) using hybrid molecular beam epitaxy (MBE) approach (details described elsewhere[25–28]). Sr was supplied using a conventional solid source effusion cell, while Ti was supplied with a gas delivery system using the metalorganic precursor titanium tetraisopropoxide (TTIP). Additional oxygen was supplied using a rf plasma source (Mantis, UK) to ensure oxygen stoichiometry. All the films were grown within the optimized growth window condition[25] and at a fixed substrate temperature of 900°C (read by thermocouple). Control undoped STO films were insulating, suggesting no measurable conduction due to oxygen vacancies. Reflection high-energy electron diffraction (RHEED) was employed to monitor the growth *in-situ*, and high-resolution X-ray diffraction (XRD) was used to determine the phase purity and cation stoichiometry *ex-situ* (Fig. 1a). We studied fourteen doped samples, with carrier densities spanning from $5.0\times10^{17}$ to $1.6\times10^{20}$ $cm^{-3}$. Milli-Kelvin measurements were performed in an Oxford Triton dry dilution refrigerator equipped with home-made RC and Pi thermalizing filters mounted on the mixing chamber plate. We observed superconductivity for samples in the density range $1.7\times10^{19}$ to $1.6\times10^{20}$ $cm^{-3}$. Below, we discuss in detail four of these samples, which are representative of the properties of the entire set studied (sheet carrier densities, $n_{2D}$, are reported in Table 1 alongside the volume carrier densities, $n$).

## RESULTS AND DISCUSSION

The observed normal-state resistance is described by a quadratic law, $R = R_0 + AT^2$, across the entire temperature range, with a change in the $R_0$ and $A$ parameters near $T$=180 K (Fig. 1(b)). A $T^2$ dependence of the resistance at low $T$ is typically attributed to Fermi liquid behavior, including in the case of STO[29]. However, unlike in typical metals, where the Fermi temperature ($T_F$) is ~$10^4$ K, in STO it can be as low as ~10 K due to its exceptionally large dielectric constant (in excess of 20,000 at low $T$). This means that the Fermi-liquid approximation, which requires $T \ll T_F$, may not generally hold for STO. Intriguingly, we find that the $T^2$ dependence extends to temperatures above $T_F$ and persists up to room temperature. At these elevated temperatures, the system transitions from Fermi liquid to a Boltzmann gas and hence can no longer be adequately described by Fermi liquid theory. Hence it is likely that the $T^2$ behavior, at least near and above $T_F$, has another root and may therefore not be a robust indicator of a Fermi liquid, as also recently pointed out in Refs.[30–32].

As our samples are cooled further, they become superconducting at temperatures ranging between 90 mK and 200 mK. These values of $T_c$ are in the same overall range as reported by previous studies on STO and STO-based interfaces[9,21,22,24,33,34]. Fig. 1(c) shows $R$ vs. $T$ curves as a function of magnetic field ($B$) for the sample with $n = 3.9\times10^{19}\,cm^{-3}$. Results from other samples are shown in the Supplemental Material, where we also discuss further details of the measurements[35].

**Out-of-plane critical field**

We now turn to the discussion of the temperature-dependence of the critical field, which underpins the key points of our paper. Fig. 2 displays $H_{c2}^{\perp}(T)$ for four of our samples. First, we note that all have non-negative curvature starting just below $T_c$. This behavior is not expected from a typical single-band BCS superconductor, and is a hallmark of multi-band superconductivity[3,5]. For a single-band superconductor, a negative curvature would instead be expected at all $T$, as described by the WHH model[36].

For the $n = 3.9\times10^{19}\,cm^{-3}$ sample a second remarkable feature of the data is that the value of $H_{c2}^{\perp}(T)$ measured at the lowest temperature (~0.25$T_c$) reaches almost twice the value of the Chandrasekhar-Clogston (Pauli paramagnetic) limit, $H_p \sim 1.84T_c\left(T/K\right)$. For thin film superconductors it is common to observe *in-plane* critical fields ($H_{c2}^{\parallel}$) exceeding $H_p$, due to the geometrical enhancement of the bulk critical field, combined with the presence of SO scattering [37,38]. However, the *out-of-plane* critical field, $H_{c2}^{\perp}(0)$ of thin films, just like $H_{c2}$ for bulk samples, is typically much less than $H_p$ in most materials because orbital pair-breaking effects are strong in this configuration. Exceptions are certain bulk organic and heavy-fermion superconductors[39,40]. In the former case, this has been attributed to a field-dependent dimensionality crossover coupled with the presence of SO scattering, while in the latter the very large effective electron mass contributes to suppressing orbital effects. Moreover, bulk critical fields as high as $H_p$ (i.e. $H_{c2}(0) \le H_p$) have been reported in iron-pnictide superconductors[6,41]. However, for STO the observation of $H_{c2}^{\perp}(0) > H_p$ is quite striking. Below, we show that $H_{c2}^{\perp}(T) > H_P$ for our sample is the consequence of the enhancement of the orbital critical field by multi-band superconductivity, coupled with the presence of strong SO scattering.

In Fig. 2 we plot our data against the single-band WHH expression in order to contrast our observations with the expectations for a single-band superconductor. It is evident that despite taking into account SO scattering, single-band superconductivity does not capture the salient features of the data, including the non-negative curvature and the large $H_{c2}^{\perp}(0)$. In single-band superconductors the slope of $H_{c2}(T)$ near $T_c$ determines the maximum value of the critical field, $H_{c2}(0)$, via the simple relation $H_{c2}(0) = -0.69 T_c \, dH_{c2}/dT\big|_{T_c}$, which can be rewritten as $H_{c2}(0) \propto \frac{2\pi c k_B T_c}{hDe}$, where $D$ is the diffusivity. This shows that the important material parameters which determine the critical field are $T_c$ and $D$. A larger $T_c$ and a smaller $D$ will result in larger critical field. Both $T_c$ and $D$ depend on the density: $T_c$ is a non-monotonic function of n (superconducting domes), while D generally decreases with increasing density.

The same considerations apply to the two-band case. However, in contrast, for two-band superconductors a large enhancement of $H_{c2}$ beyond the single-band prediction occurs if the diffusivity of one of the bands is small. This is possible because $H_{c2}(0)$ is no longer simply determined by $dH_{c2}/dT\big|_{T_c}$. A theoretical picture of two-band superconductivity was first developed by Suhl[42] and further extended by Gurevich to calculate critical fields[3,20]. For two-band superconductors subject to orbital effects alone, the Gurevich model predicts that $H_{c2}(0)$ is given by[3]:

$$H_{c2}(0) = \frac{c k_B}{\pi \gamma e D_1} e^{-(\lambda_0 - \lambda_-)/2w}, \qquad \text{for } D_1 \ll D_2 e^{-\lambda_0/2w} \qquad (1)$$

Here $D_1$ and $D_2$ correspond to the diffusion constants of the two bands (we denote the lower, heavier band as band 1), $\lambda_- = \lambda_{11} - \lambda_{22}$, $\lambda_0 = \sqrt{(\lambda_{11} - \lambda_{22})^2 + 4\lambda_{12}\lambda_{21}}$, and $w = \lambda_{11}\lambda_{22} - \lambda_{12}\lambda_{21}$, with $\lambda_{ij}$ as the intra- and inter-band coupling constants. $\ln(\gamma) = -0.577$ is the Euler constant, and $c$, $k_B$ and $e$ are the speed of light, Boltzmann constant and elementary charge, respectively. Importantly, $H_{c2}(0)$ is determined by the lower of the two diffusivities. This unique feature of two-band superconductors allows much enhanced $H_{c2}(0)$ values with respect to the single-band case. For $D_1/D_2 \ll 1$ or $D_1/D_2 \gg 1$ a pronounced positive curvature and significant enhancement of $H_{c2}$ compared to the single band case are expected. On the other hand, for $D_1/D_2 \cong 1$, the critical field behavior resembles qualitatively that of single-band superconductors. Increasing disorder tends to affect disproportionately one of the bands, leading to more pronounced two-band effects, such as $H_{c2}(T)$ positive curvature[18,20]. In STO, three bands become occupied for carrier densities above $\sim 2$ to $3 \times 10^{19}$ $cm^{-3}$ [12], however, a two-band model is expected to provide a good description because the difference in energy between the lower two bands is between 3-5 meV, while that between the middle and upper bands is about 12-30meV. As a result, inter-band scattering can be significant for the lower two bands and therefore these two bands can be approximated as a single band, especially when all three bands are occupied, as supported by theoretical modeling[49].

We note that earlier examples of out-of-plane critical fields exceeding the Pauli-limit can be found in the STO literature[22,23]. These references show data where $H_{c2}$ is up to ~1.6 $H_p$, although the effect was not noticed or discussed in those works. This suggests that this unusual and interesting effect could be a general property of doped STO under the right experimental conditions, although what these conditions are is at present not known.

We next proceed to analyzing the *T*-dependence of $H_{c2}^{\perp}$ using the prediction of the Gurevich two-band model, noting that for a thin film $H_{c2}^{\perp}$ is effectively $H_{c2}$ of the bulk material. The Gurevich model has been previously used to fit $H_{c2}(T)$ data on two-band superconductors such as $MgB_2$ and Fe-pnictides with good quantitative agreement[5,43]:

$$a_0\left(\ln t + U(h)\right)\left(\ln t + U(\eta h)\right) + a_1\left(\ln t + U(h)\right) + a_2\left(\ln t + U(\eta h)\right) = 0 \qquad (2)$$

Here, $U(x) = \psi(1/2 + x) - \psi(1/2)$, where $\psi$ is the di-gamma function. $a_{0,1,2}$ are determined by the band coupling constants $\lambda_{ij}$ as described in Ref.[3], $h = D_2 eH/2\pi c k_B T$ (we use cgs units), $t = T/T_c$, and $\eta = D_1/D_2$. Note that in the limit of $\eta = 1$ eqn. (2) reduces to the de Gennes-Maki equation, $\ln t + U(h) = 0$. To determine $a_{0,1,2}$ we use the coupling constants obtained in [44], $\lambda_{11} = \lambda_{22} = 0.14$ and $\lambda_{12} = \lambda_{21} = 0.02$, indicative of a dominant intra-band coupling. We note that our data is not consistent with the other published set of estimates for $\lambda_{ij}$ [45]. This provides experimental constraints on the values of these constants[35]. Furthermore, the data is inconsistent with dominant inter-band coupling, $\lambda_{12}\lambda_{21} >> \lambda_{11}\lambda_{22}$, or $\lambda_{12}\lambda_{21} \approx \lambda_{11}\lambda_{22}$, suggesting that intra-band coupling is dominant[35].

Having thus fixed the $\lambda_{ij}$ using the values from Ref. [44], we are left with only two adjustable parameters, $D_1$ and $D_2$. The slope of $H_{c2}$ vs. $T$ near $T_c$ is determined by the sum of $D_1$ and $D_2$. Hence, by fixing this sum there remains only one free parameter for the rest of the data. Although the Gurevich two-band orbital model does not account for paramagnetic and SO scattering effects, we find a good agreement with our $H_{c2}^{\perp}(T)$ data (Fig. 2). The best fit values of $D_1$ and $D_2$ are listed in Table 1. Note that the ratio $\eta=D_1/D_2$ varies between ~0.02 and 0.3, indicating a large difference between the diffusivities of the two bands, as well as variations between samples with different carrier concentrations and thicknesses. We find that the sum of $D_1$ and $D_2$ is comparable to the value of $D$ extracted from the normal state properties. Taking the sample with $n_{3D} = 3.9 \times 10^{19}\ cm^{-3}$ as example, we estimate the diffusion constant obtained from the normal state resistivity to be $D \approx 2\ \mathrm{cm}^2/\mathrm{s}$. For comparison, the fits to the superconducting critical field data yield $D_1 + D_2 \approx 3.06\ \mathrm{cm}^2/\mathrm{s}$, which is close to the value of D. As a broader point, we note that order of magnitude discrepancies can be observed between the diffusivities obtained from normal state and superconductivity measurements in doped STO[48], which could be due to different bands playing a dominant role in the normal and superconducting states.

Importantly, the model clearly displays a non-negative curvature, in agreement with the data and in contrast with the single-band WHH model. For completeness, we note that a positive curvature of $H_{c2}(T)$ has been attributed to melting of the vortex lattice in high-$T_c$ cuprates and has also been theoretically associated to paramagnetic impurities[46,47]. However, since STO is a low-$T_c$ superconductor, vortex fluctuations are negligible and so we can rule out the melting of a vortex lattice in our case. Moreover, the concentrations of $Nd^{3+}$, x=0.1% to 0.3% are unlikely to be sufficient to lead to any spin-spin correlations here.

**In-plane critical field and our extended model**

Having shown that the $H_{c2}^{\perp}(T)$ data reveals the presence of two-band superconductivity in our samples, we now turn to a discussion of $H_{c2}^{\parallel}(T)$ (Fig. 3), which reveals additional physics. We first proceed with an analysis based on the Gurevich two-band model for $H_{c2}^{\parallel}(T)$. We use $h=1/6D_2e^2H^2d^2/\hbar c^2$, consistent with Ref. [20]. In Fig. 3 we show the prediction of this model using the parameter values obtained above for $H_{c2}^{\perp}(T)$, with only one adjustable parameter, $d$, the effective film thickness. We find that the model describes the data well near $T_c$, but substantially overestimates the low-$T$ data for two of the four samples (best fit values are shown in Table 1). We argue that this overestimation stems from ignoring paramagnetic effects. Moreover, since experimentally $H_{c2}^{\parallel}(0) > H_p$ for all four samples, SO scattering effects must also be taken into account. This motivates the need for a more comprehensive two-band model, which includes orbital, paramagnetic and SO scattering effects, in order to account for the totality of our observations. To our knowledge no such models have been developed.

Here we propose an extension of the Gurevich two-band model, which includes all these effects. The generalization of the single-band critical field equation to the case of two-band systems is generally straightforward. In the limit of negligible inter-band scattering, the procedure entails solving two single band equations independently and casting these into a two by two matrix equation. The diagonal matrix elements contain the intra-band coupling constants, while off-diagonal elements contain the inter-band coupling constants. This yields eqn. (2), with $U(h)$ and $U(\eta h)$ replaced with the appropriate form for the single-band model. All the existing two-band superconductor critical field equations can be understood in this fashion[3,20]. In the case of a two-band superconductor in the dirty limit with orbital, paramagnetic, and SO effects, the relevant single-band model is the WHH model, and hence extending the two-band model to include paramagnetic and SO scattering effects involves replacing $U(\eta h)$ and $U(h)$ in eqn. (2) with the WHH expressions, $U^*(\eta \bar{h}, \alpha_1, \lambda_{so,1})$ and $U^*(\bar{h}, \alpha_2, \lambda_{so,2})$, as defined below (additional details are provided in the Supplemental Material [35]):

$$U^*(\bar{h}, \alpha_i, \lambda_{so,i}) = \left(\frac{1}{2} + \frac{i\lambda_{so,i}}{4\gamma_i}\right)\psi\left(\frac{1}{2} + \frac{\bar{h} + \frac{1}{2}\lambda_{so,i} + i\gamma_i}{2t}\right) + \left(\frac{1}{2} - \frac{i\lambda_{so,i}}{4\gamma_i}\right)\psi\left(\frac{1}{2} + \frac{\bar{h} + \frac{1}{2}\lambda_{so,i} - i\gamma_i}{2t}\right) - \psi\left(\frac{1}{2}\right) \quad (3)$$

As in the original WHH paper, $\lambda_{so,i} = 2\hbar/3\pi k_B T_c \tau_{so,i}$, where $\tau_{so,i}$ is the spin-orbital scattering time, $\bar{h} = \frac{D_2 eH}{\pi c k_B T_c}$, and $\gamma_i = \sqrt{\left(\alpha \bar{h}\right)^2 - \left(\frac{1}{2}\lambda_{so,i}\right)^2}$, where $\alpha_i = \frac{\hbar}{2mD_i}$. Here $\lambda_{so,i}$ represents the characteristic spin-orbital coupling energy for each band normalized by the condensation energy, and $\gamma_i$ includes the paramagnetic (Zeeman) term. Note that for small applied fields, $\frac{\mu_B H}{\pi k_B T_c} \ll \frac{1}{2}\lambda_{so,i}$, where orbital effects are dominant, we recover the orbital-only critical field model given by eqn. (2)[3], while in the absence of SO scattering ($\lambda_{so,i} = 0$), this reduces to the Gurevich model with paramagnetic effects[20], in the limit of vanishing inter-band scattering.

To assess our extended model, we fit both the $H_{c2}^{\perp}(T)$ and $H_{c2}^{\parallel}(T)$ data using the same literature values of the coupling constants $a_{0,1,2}$ as used above for the orbital-only model and display the results as the green solid lines in Fig. 2 and Fig. 3. There are five adjustable parameters: $D_1$, $D_2$, $\lambda_{so,1}$, $\lambda_{so,2}$ and $d$. However, physical insight about the role and type of SO scattering (detailed below) helps reduce their effective number. Moreover, the fact that the parameters must satisfy both the $H_{c2}^{\perp}(T)$ and the $H_{c2}^{\parallel}(T)$ data simultaneously places a further constraint on the model. To fit to both data sets we use an iterative approach. We begin by obtaining an initial best fit to the $H_{c2}^{\perp}(T)$ data, which we find to be relatively insensitive to the SO scattering strength, so that the initial fit is essentially based on only $D_1$ and $D_2$. Using these initial values for $D_1$ and $D_2$, we then fit to the $H_{c2}^{\parallel}(T)$ data by varying $\lambda_{so,1}$, $\lambda_{so,2}$ and $d$. We note that the $H_{c2}^{\parallel}(T)$ data is primarily sensitive to the value of $d$ for $H < H_p$ because SO scattering effects are weak for this range of fields. Conversely, it is most sensitive to $\lambda_{so,1}$ and $\lambda_{so,2}$ for $H > H_p$, where SO and paramagnetic effects play the biggest role. The iterative process is then repeated, using the values of $\lambda_{so,1}$, $\lambda_{so,2}$ and $d$ so obtained to yield a better fit to $H_{c2}^{\perp}(T)$ at the next iteration. We repeat the iterations until the fits to both data sets converge and the parameters no longer change appreciably. To further constrain $\lambda_{so,1}$ and $\lambda_{so,2}$ and to shed light on the origin of the SO scattering, we consider two commonly occurring SO scattering mechanisms: Elliot-Yafet[50] (EY) and Dyakonov-Perel[51] (DP). For EY, the transport lifetime ($\tau_p$) is proportional to the spin lifetime, $\tau_{so} \propto \tau_p$. The opposite is true for the DP mechanism: $\tau_{so} \propto 1/\tau_p$. Using $v_{F2}^2 \simeq 10 v_{F1}^2$ [10,17] and $D_i = v_{F,i}^2 \tau_{p,i}/3$ we find that $\lambda_{so,1}/\lambda_{so,2} \simeq (1/10)\, D_2/D_1 = 1/(10\eta)$ for EY and $\lambda_{so,1}/\lambda_{so,2} = 10\eta$ for DP SO scattering. This allows the SO scattering to be described by only one parameter instead of two. To determine which scattering mechanism is likely dominant, we perform the fits with each type on the $n = 3.9\times10^{19} cm^{-3}$ sample. Assuming that the DP mechanism dominates, we obtain a good fit to the data with $D_1$ ~ 0.04 cm$^2$/s, $D_2$ ~ 3 cm$^2$/s, $\lambda_{so,1} \sim 2$ and $\lambda_{so,2} \sim 20$. The value $\lambda_{so,2} \sim 20$ corresponds to a spin-orbit energy $h/k_B\tau_{so,2} \sim 120K$, which is rather large, suggesting that the DP mechanism is in fact not the main source of SO scattering. Indeed, Dresselhaus-type DP SO scattering is precluded by the fact that the STO unit cell is centrosymmetric, although Rashba-type SO coupling and DP SO scattering occur in certain STO-based surfaces and interfaces due to structural inversion asymmetry[52–56]. On the other hand, fitting under the assumption that EY dominates, we obtain $D_1$ ~ 0.04 cm$^2$/s, $D_2$ ~ 3 cm$^2$/s, $\lambda_{so,1} \sim 4$ and $\lambda_{so,2} \sim 0.4$, with $d$ ~ 8 nm. In this case, the maximum SO scattering energy ($h/k_B\tau_{so,1}$) corresponds to ~20K. The dominance of EY over DP is consistent with the absence of inversion symmetry breaking in STO. (The fit parameters for the remaining three samples are shown in Table 1, for dominant EY SO scattering.) Interestingly, the obtained SO scattering energies $h/\tau_{so,i}$ of up to $\sim 2$ meV are significantly larger than the condensation energy of ~30 μeV. We note that in STO, a third type of SO mechanism, intrinsic SO coupling, could also play a role[37], however there is currently relatively little experimental understanding of the effects of this mechanism on superconductivity. Another outcome of the fits is that the extracted film thicknesses ($d$) are in the 8 to 25 nm range (Table 1), which is close to the values obtained from the Gurevich orbital-only model, but much less than the nominal thickness of the $Nd_xSr_{1-x}TiO_3$ layer, which ranges between 60 and 230 nm. This indicates that the samples behave effectively as two-dimensional superconductors, as corroborated by the angular-dependence of $H_{c2}$, described next. The extracted thickness of the superconducting layer does not necessarily correspond to the thickness in which mobile electrons are present. Given the strong carrier density dependence of superconductivity in STO, if the density varies somewhat across the film thickness, this can make a thin slice of the film dominate in the superconducting measurements, even though most of the film may contribute to the normal state transport. This can arise, e.g. due to surface charge depletion, a large effect in STO due to its exceptionally large dielectric constant[57]. In addition, STO is well-known for its tetragonal domain structure, which could also play a role[58,59].

**Angular dependence of the critical field**

We conclude our analysis with a discussion of the angular dependence of $H_{c2}$, which is typically used as test of the effective dimensionality of a superconductor. Fig. 4 shows plots of $H_{c2}$ vs. $\theta$, the angle between the applied field and the normal to the sample. To describe the angular dependence of $H_{c2}$ between the out-of-plane and in-plane limits, we use $h(\theta)=\frac{D_2}{2\pi c k_B T_c}\left(2eH\left|\cos(\theta)\right|+\frac{1}{3\hbar c}\left[deH\sin(\theta)\right]^2\right)$. We find a good agreement of our model with data over a wide range of angles, *with no fitting parameters*, using the values extracted previously from fitting $H_{c2}^{\perp}(T)$ and $H_{c2}^{\parallel}(T)$. For comparison, we also fitted the data to the single-band 2D (Tinkham) and anisotropic 3D (Ginzburg-Landau) models, which have two adjustable parameters, $H_{c2}^{\perp}$ and $H_{c2}^{\parallel}$. Although neither of these models takes paramagnetic or SO scattering effects into account, the 2D Tinkham model is frequently used to fit angular dependent data for thin films, even where the in-plane critical field is significantly larger than $H_p$, often with good agreement to the data[38,60]. We find that the 2D Tinkham model yields a satisfactory fit, however the full model including paramagnetic and SO scattering effects provides an overall better fit across the samples studied, despite having no free parameters vs. two for the Tinkham model. In contrast, the 3D anisotropic Ginzburg-Landau model produces a rounded peak (insets of Fig. 4), in clear deviation from our data, confirming the 2D nature of the observed superconductivity. A possible reason to consider for the observed 2D behavior could be the onset of surface superconductivity, that is the presence of a surface superconducting layer with critical field $H_{c3}$

~1.7 $H_{c2}$, which can occur in samples with very clean surfaces when a magnetic field is applied parallel to the surface[61]. Under this scenario, the measured parallel critical fields would be $H_{c3}^{\parallel}$ rather than $H_{c2}^{\parallel}$. However, since this effect is unstable for perpendicular applied fields, it is unlikely to contribute to the observed $H_{c2}^{\perp}$ (T) behavior. STO-vacuum surfaces have also been shown to host a natural two-dimensional electron gas localized to within a few unit cells of the surface in cleaved samples[53,55], presumably due to oxygen vacancies[53,62]. This mechanism is also unlikely to play a role in our samples, which were grown in an oxygen-rich environment and are insulating when not doped with Nd.

## CONCLUSIONS

In conclusion, we showed that the temperature-dependence of the upper critical field can be a powerful and sensitive approach for investigating the interplay between two-band superconductivity and spin-orbit coupling in the dirty limit. In particular, we proposed an approach for analyzing this temperature-dependence when orbital pair-breaking effects coexist with paramagnetic and spin-orbital effects. We found a very good agreement between our model and our data on Nd-doped $SrTiO_3$ thin films, which suggests that this model could be used successfully for other materials systems in which these mechanisms are at play, including STO thin films with various dopants, STO-based interfaces and others. Interestingly, we find that under the right circumstances the zero-temperature critical field of such systems can exceed the Pauli limit even for magnetic fields applied perpendicular to the film plane, as confirmed experimentally.

**Acknowledgements:**

We thank Rafael Fernandes, Allen Goldman, Ali Yazdani and Thais Trevisan for valuable discussions, and Koustav Ganguly for technical support. This work was supported primarily by the National Science Foundation (NSF) Materials Research Science and Engineering Center at the University of Minnesota under Award No. DMR-1420013. Film growth and structural characterizations were funded by the U.S. Department of Energy through the University of Minnesota Center for Quantum Materials, under Grant No. DE-SC-0016371. Portions of this work were conducted in the Minnesota Nano Center, which is supported by the National Science Foundation through the National Nano Coordinated Infrastructure Network (NNCI) under Award Number ECCS-1542202. Sample structural characterization was carried out at the University of Minnesota Characterization Facility, which receives partial support from NSF through the MRSEC program. VSP acknowledges funding from the Alfred P. Sloan foundation.

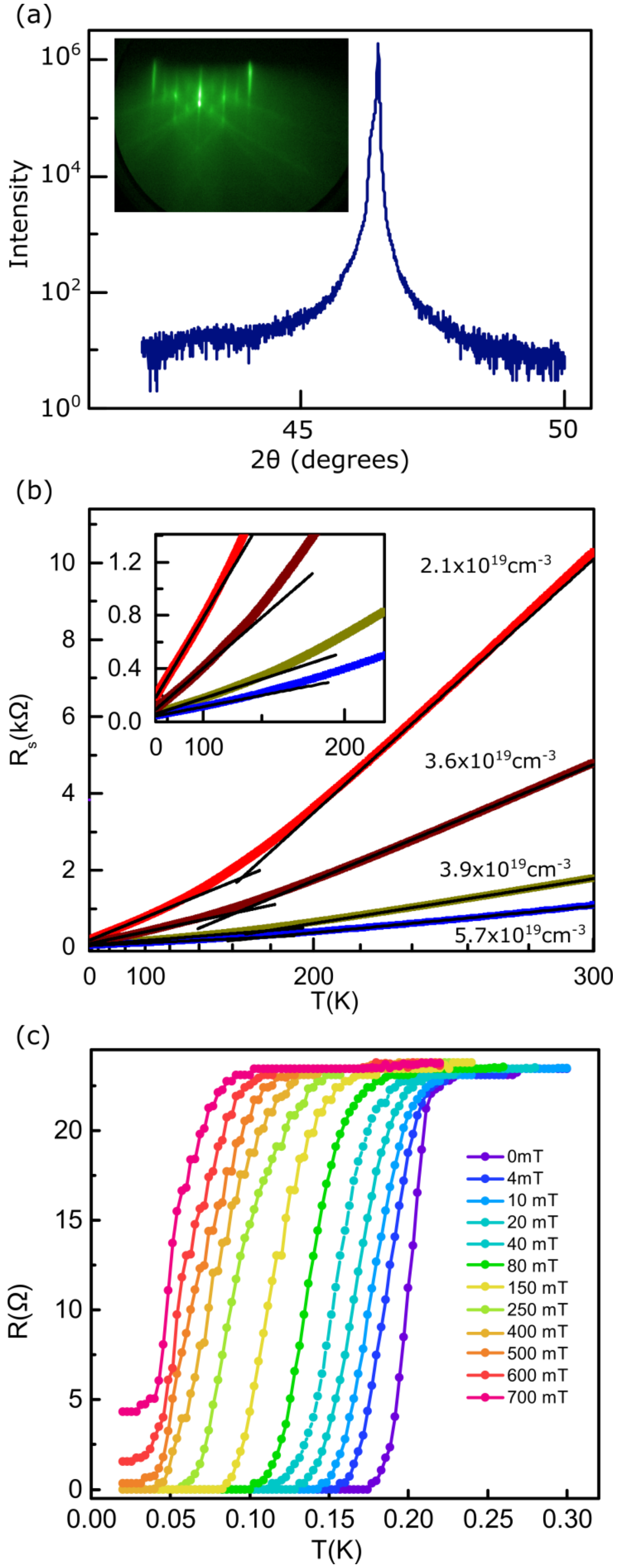

**Fig. 1:** (a) XRD and RHEED (inset) data on sample with $n = 3\times10^{19} cm^{-3}$ indicating an epitaxial, single-crystalline, smooth film. (b) Normal state sheet resistance ($R_s$) as function of temperature ($T$), for four of our samples, plotted on $T^2$ scale, showing quadratic dependence of $R$ on $T$ from 2K to 300 K, with a crossover in the $T^2$ prefactor around 180K. From top to bottom, the samples have $n = 2.1,\ 3.6,\ 3.9$ and $5.7\times10^{19} cm^{-3}$ respectively. The thicknesses of the doped STO layer are 60 nm, 60 nm, 230 nm and 220 nm respectively (see Table 1). For reference, $T_F$~150K for the sample with $n = 3.9\times10^{19} cm^{-3}$, in the parabolic band approximation. Inset: Zoom-in of the data for T < 200K. (c) Resistance ($R$) as function temperature ($T$) at different out of plane fields for the sample with $n = 3.9\times10^{19} cm^{-3}$, showing the superconducting transition. Note that the apparent saturation of resistance below ~40 mK is due to reaching the minimum temperature to which the sample can be cooled, rather than actual saturation of the sample resistance.

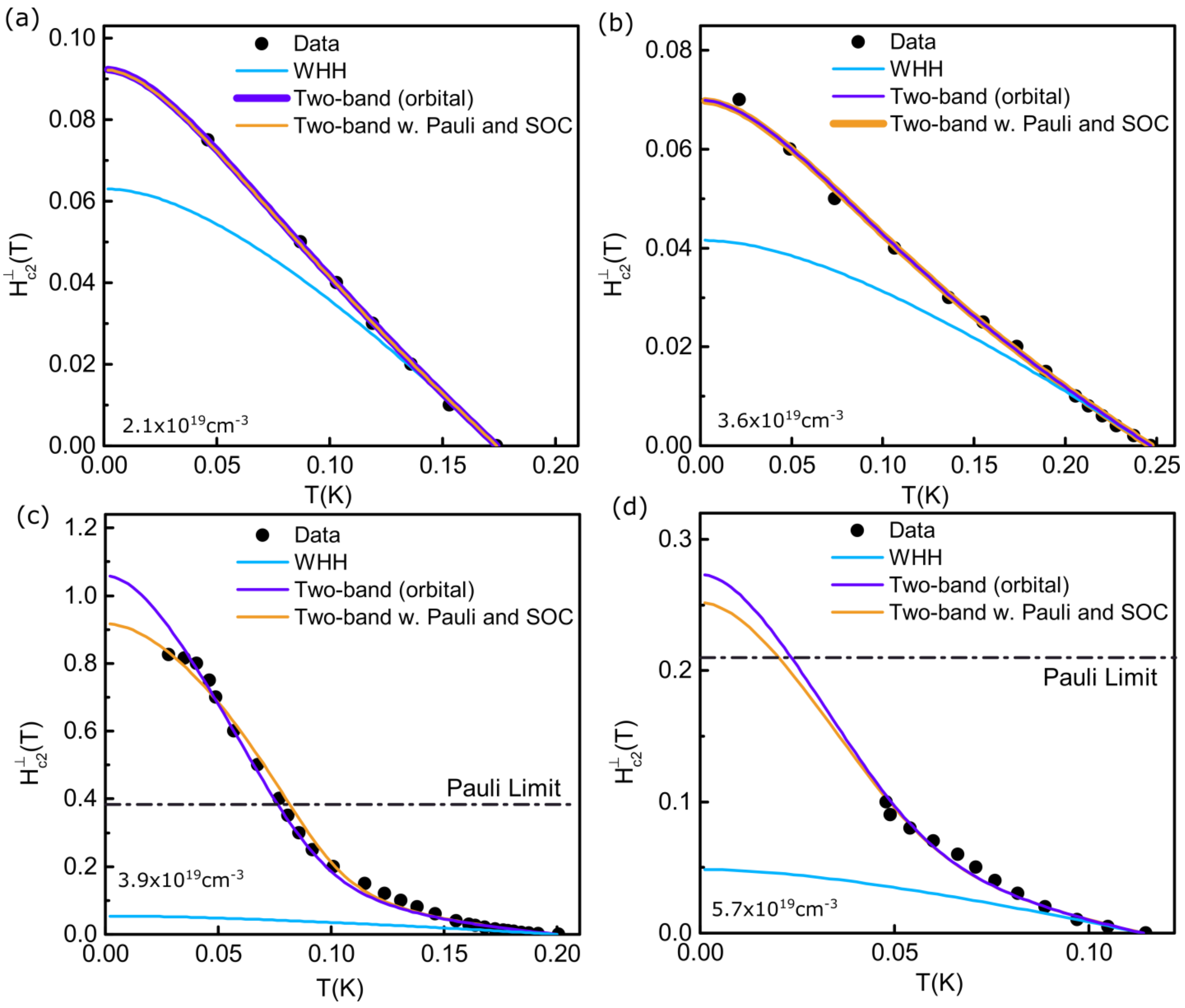

**Fig-2:** (a-d) Out-of-plane upper critical field vs. temperature for four samples, with $n = 2.1,\ 3.6,\ 3.9$ and $5.7\times10^{19}\,cm^{-3}$ respectively. Solid curves correspond to fits using single-band WHH, two-band model with orbital effects only and two-band model with orbital, paramagnetic and SO scattering effects. Note the different temperature ranges in the four plots, due to the different $T_c$s of the samples. For the two samples in (c) and (d) the results of the Gurevich two-band model overlap with those of the two-band model with Pauli paramagnetic and SO scattering effects.

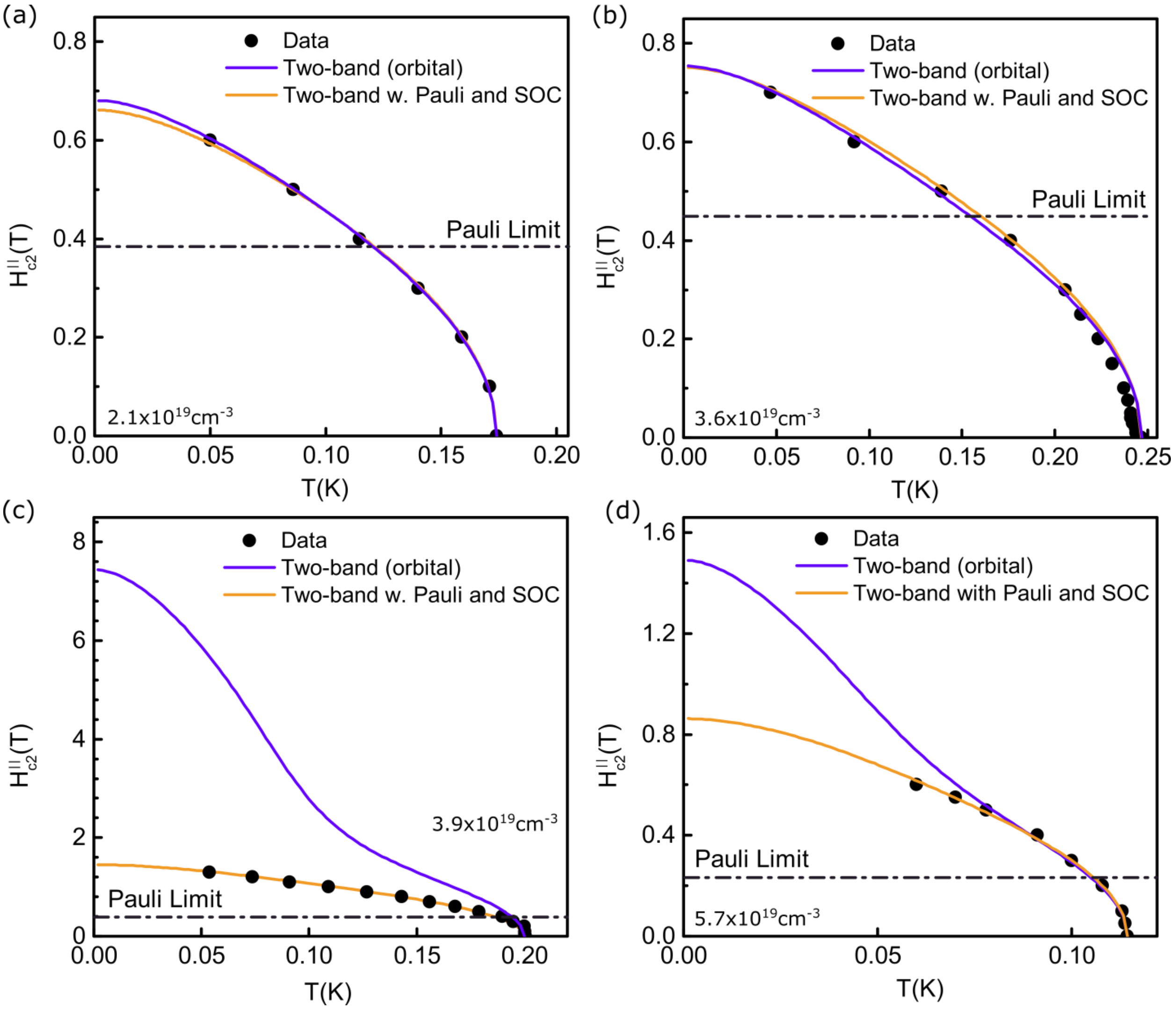

**Fig. 3**: (a-d) In-plane upper critical field vs. temperature for four samples, with $n = 2.1,\ 3.6,\ 3.9$ and $5.7\times10^{19}\,cm^{-3}$ respectively. Solid curves correspond to fits using the two-band model with orbital effects only and the two-band model with orbital, paramagnetic and SO scattering effects.

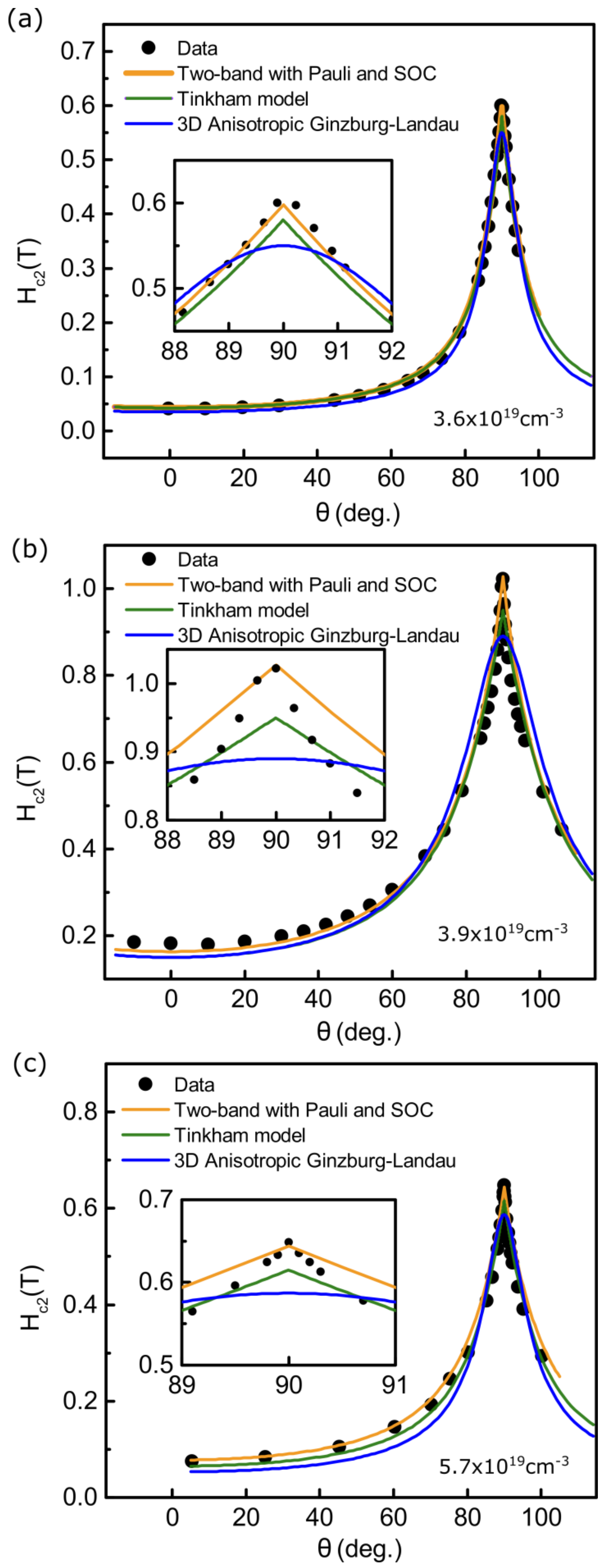

**Fig. 4:** (a-c) Upper critical field as function of angle measured from the normal to the sample plane for three samples, with $n = 3.6,\ 3.9,$ and $5.7\times10^{19}cm^{-3}$ respectively, at temperatures ~100 mK, 108 mK and 60 mK, respectively. Insets show zoom-in around $90^0$ (field nearly aligned to sample plane). The different curves correspond to fits using the two-band model with orbital, paramagnetic and SO scattering effects, the 2D Tinkham model and the anisotropic Ginzburg-Landau model.

| | $n = 2.1\times10^{19} cm^{-3}$ $n_{2D} = 1.3\times10^{14} cm^{-2}$ (60-nm doped STO) | $n = 3.6\times10^{19} cm^{-3}$ $n_{2D} = 2.2\times10^{14} cm^{-2}$ (60-nm doped STO) | $n = 3.9\times10^{19} cm^{-3}$ $n_{2D} = 9.0\times10^{14} cm^{-2}$ (230-nm doped STO) | $n = 5.7\times10^{19} cm^{-3}$ $n_{2D} = 1.3\times10^{15} cm^{-2}$ (220-nm doped STO) |
|---|---|---|---|---|
| **WHH model** | D=2.1 $cm^2/s$ | D=4.5 $cm^2/s$ | D=1.5 $cm^2/s$ | D=1.6 $cm^2/s$ |
| **Two-band (orbital) model** | $D_1$=0.9 $cm^2/s$<br>$D_2$=3.3 $cm^2/s$<br>d=28 nm | $D_1$=1.6 $cm^2/s$<br>$D_2$=7.4 $cm^2/s$<br>d=22 nm | $D_1$=0.06 $cm^2/s$<br>$D_2$=3.0 $cm^2/s$<br>d=8 nm | $D_1$=0.16 $cm^2/s$<br>$D_2$=3.1 $cm^2/s$<br>d=22 nm |
| **Two-band model w. Pauli and SO scattering effects** | $D_1$=0.90 $cm^2/s$<br>$D_2$=3.3 $cm^2/s$<br>$\lambda_{so,1}$=2.0<br>$\lambda_{so,2}$=5.5<br>d=26 nm | $D_1$=1.6 $cm^2/s$<br>$D_2$=7.4 $cm^2/s$<br>$\lambda_{so,1}$=2.5<br>$\lambda_{so,2}$=5.5<br>d=20 nm | $D_1$=0.036 $cm^2/s$<br>$D_2$=3.0 $cm^2/s$<br>$\lambda_{so,1}$=3.5<br>$\lambda_{so,2}$=0.4<br>d=8 nm | $D_1$=0.16 $cm^2/s$<br>$D_2$=3.1 $cm^2/s$<br>$\lambda_{so,1}$=1.2<br>$\lambda_{so,2}$=0.6<br>d=17 nm |

**Table 1:** Values of fit parameters for the models discussed in the text.

# SUPPLEMENTAL MATERIAL

### Effects of paramagnetic pair-breaking and spin-orbital coupling on multi-band superconductivity

Yilikal Ayino[1], Jin Yue[2], Tianqi Wang[2], Bharat Jalan[2] and Vlad S. Pribiag[1*]

[1]School of Physics and Astronomy, University of Minnesota
[2]Department of Chemical Engineering and Materials Science, University of Minnesota
[*]Corresponding author: vpribiag@umn.edu

## I. Additional details on the extended model for two-band superconductivity including Pauli pair-breaking and SO scattering

Following Refs. [3] and [20] of the main text, we write the equations for the two gaps, $\Delta_1$ and $\Delta_2$, in terms of the $l - U^*$ function, which describes the dependence of the critical field on the relevant parameters ($l = ln(2\gamma\omega_D / \pi T)$, where $\gamma$ is the Euler constant and $\omega_D$ is the Debye frequency). The first term in the coupled equations below represents the intra-band coupling (parametrized by $\lambda_{11}$ and $\lambda_{22}$ for bands 1 and 2, respectively), and the second the inter-band coupling (parametrized by $\lambda_{12}$ and $\lambda_{21}$). This approach is the same as in Refs. [3] and [20], but now $U^*$ is allowed to depend on the SO scattering parameters, $\lambda_{so,1}$ and $\lambda_{so,2}$, and also accounts for Pauli pair-breaking via the $\alpha$ parameter.

$$\Delta_1 = \lambda_{11}[l - U^*(h^*, \alpha, \lambda_{so,1})]\Delta_1 + \lambda_{12}[l - U^*(\eta h^*, \eta\alpha, \lambda_{so,2})]\Delta_2$$

$$\Delta_2 = \lambda_{22}[l - U^*(\eta h^*, \eta\alpha, \lambda_{so,2})]\Delta_2 + \lambda_{21}[l - U^*(h^*, \alpha, \lambda_{so,1})]\Delta_1$$

This can be written in matrix form as:

$$A\begin{pmatrix}\Delta_1\\ \Delta_1\end{pmatrix} = 0$$

with $$A = \begin{pmatrix}\left(l - U^*(h^*, \alpha, \lambda_{so,1})\right)\lambda_{11} - 1 & \left(l - U^*(\eta h^*, \eta\alpha, \lambda_{so,2})\right)\lambda_{12}\\ \left(l - U^*(h^*, \alpha, \lambda_{so,1})\right)\lambda_{21} & \left(l - U^*(\eta h^*, \eta\alpha, \lambda_{so,2})\right)\lambda_{22} - 1\end{pmatrix}$$

This equation has a non-trivial solution at $H$=$H_{c2}$, where the gaps vanish. As a result, to find the expression for $H_{c2}$ we solve for det A=0, as in Refs. [3] and [20].

For $U^*$ we substitute the form relevant for the single-band model that includes SO and Pauli pair-breaking, namely the WHH model [34]:

$$U^*(h^*,\alpha,\lambda_{so,i})=\left(\frac{1}{2}+\frac{i\lambda_{so,i}}{4\gamma}\right)\psi\left(\frac{1}{2}+\frac{h^*+\frac{1}{2}\lambda_{so,i}+i\gamma}{2t}\right)+\left(\frac{1}{2}-\frac{i\lambda_{so,i}}{4\gamma}\right)\psi\left(\frac{1}{2}+\frac{h^*+\frac{1}{2}\lambda_{so,i}-i\gamma}{2t}\right)-\psi\left(\frac{1}{2}\right)$$

where $h^* = \frac{D_1 eH}{\pi k_B T_c c}$ for perpendicular applied fields [34] and $h^* = D_2 e^2 H^2 d^2/6hc^2$ for in-plane applied fields (by analogy with [17]), and

$$\lambda_{so,i} = \frac{2\hbar}{3\pi k_B T_c \tau_{so,i}};\ \alpha = \frac{\hbar}{2mD_1}\ ;\ \gamma = \sqrt{(\alpha h^*)^2 - \left(\frac{1}{2}\lambda_{so,i}\right)^2}$$

Solving the determinant equation yields the following implicit algebraic equation for the critical $h^*$ and hence for critical field, $H_{c2}$:

$$\begin{aligned}&a_0\left(\ln t+U^*(h^*,\alpha,\lambda_{so,1})\right)\left(\ln t+U^*(\eta h^*,\eta\alpha,\lambda_{so,2})\right)+\\&a_1\left(\ln t+U^*(h^*,\alpha,\lambda_{so,1})\right)+a_2\left(\ln t+U^*(\eta h^*,\eta\alpha,\lambda_{so,2})\right)=0\end{aligned}$$

$$a_0 = \frac{2(\lambda_{11}\lambda_{22}-\lambda_{12}\lambda_{21})}{\lambda_0};\ \ a_1 = 1+\frac{\lambda_{11}-\lambda_{22}}{\lambda_0};\ a_2 = 1+\frac{\lambda_{22}-\lambda_{11}}{\lambda_0}\ \text{and}\ \lambda_0 = \sqrt{(\lambda_{22}-\lambda_{11})^2+4\lambda_{12}\lambda_{21}};$$

This implicit equation describes the temperature-dependence of $H_{c2}$, which we fit against the data in Figs. 2 and 3 of the main text.

## II. Considering the possibility of dominant *inter*-band coupling

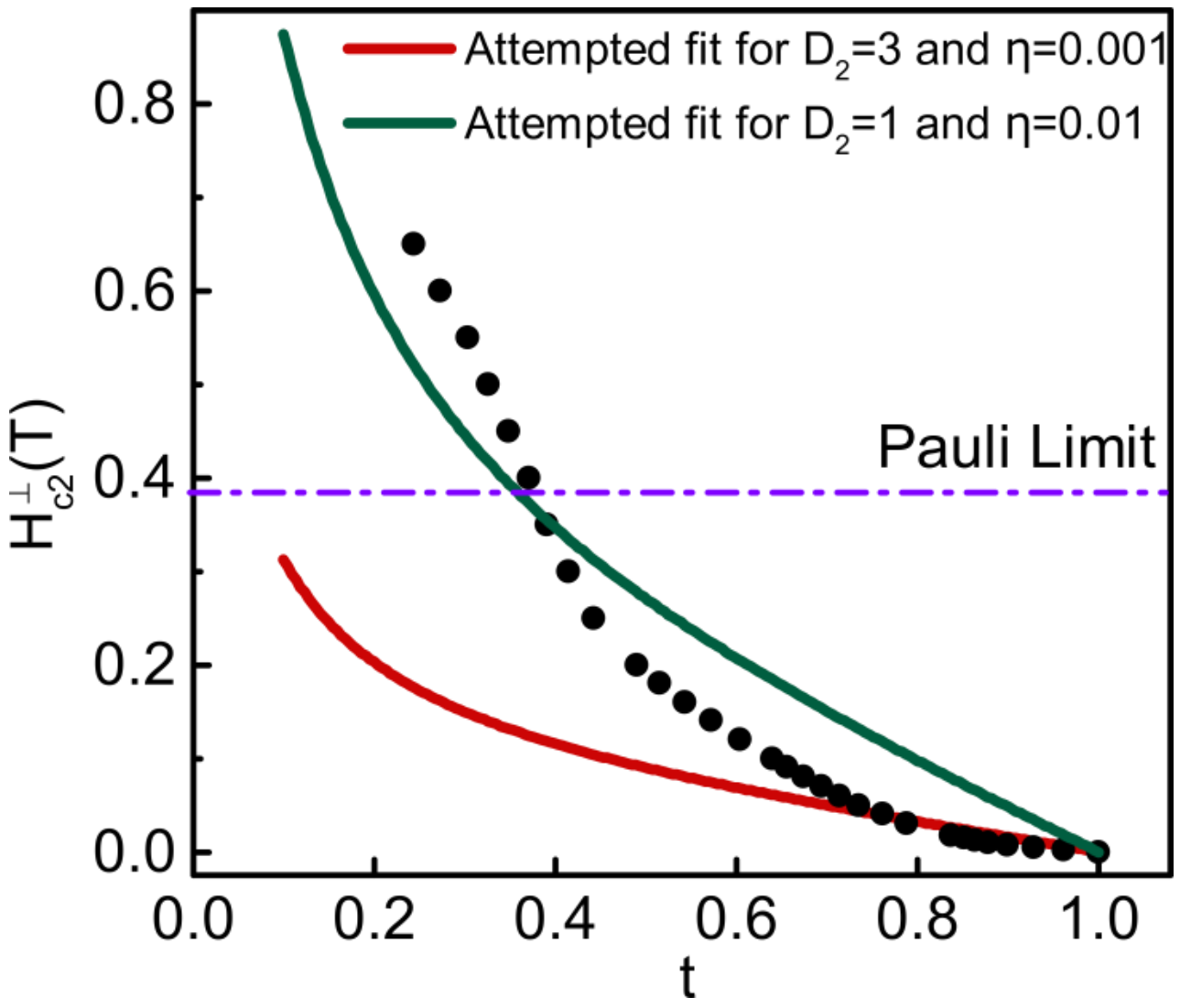

**Supplementary Fig. 1:** Out-of-plane critical field, $H_{c2}^{\perp}(T)$ for the sample with $n = 3.9\times10^{19}\,cm^{-3}$, plotted as a function of reduced temperature ($t$=$T$/$T_c$). The solid curves correspond to best obtained fits to the orbital-only equation using $\lambda_{11} = \lambda_{22} = 0.05$ and $\lambda_{12} = \lambda_{21} = 0.15$, corresponding to dominant inter-band coupling. The two fit lines (red and green) differ in the values of the diffusivities, in an attempt to capture the low-t and high-t regimes as well as possible. In both cases, a poor fit is obtained, suggesting that the data is inconsistent with dominant *inter*-band coupling. In contrast, as shown in the main paper, dominant *intra*-band coupling is in good agreement with the data.

**III. Comparing the coupling constants from Ref. [41] with those from Ref. [42]**

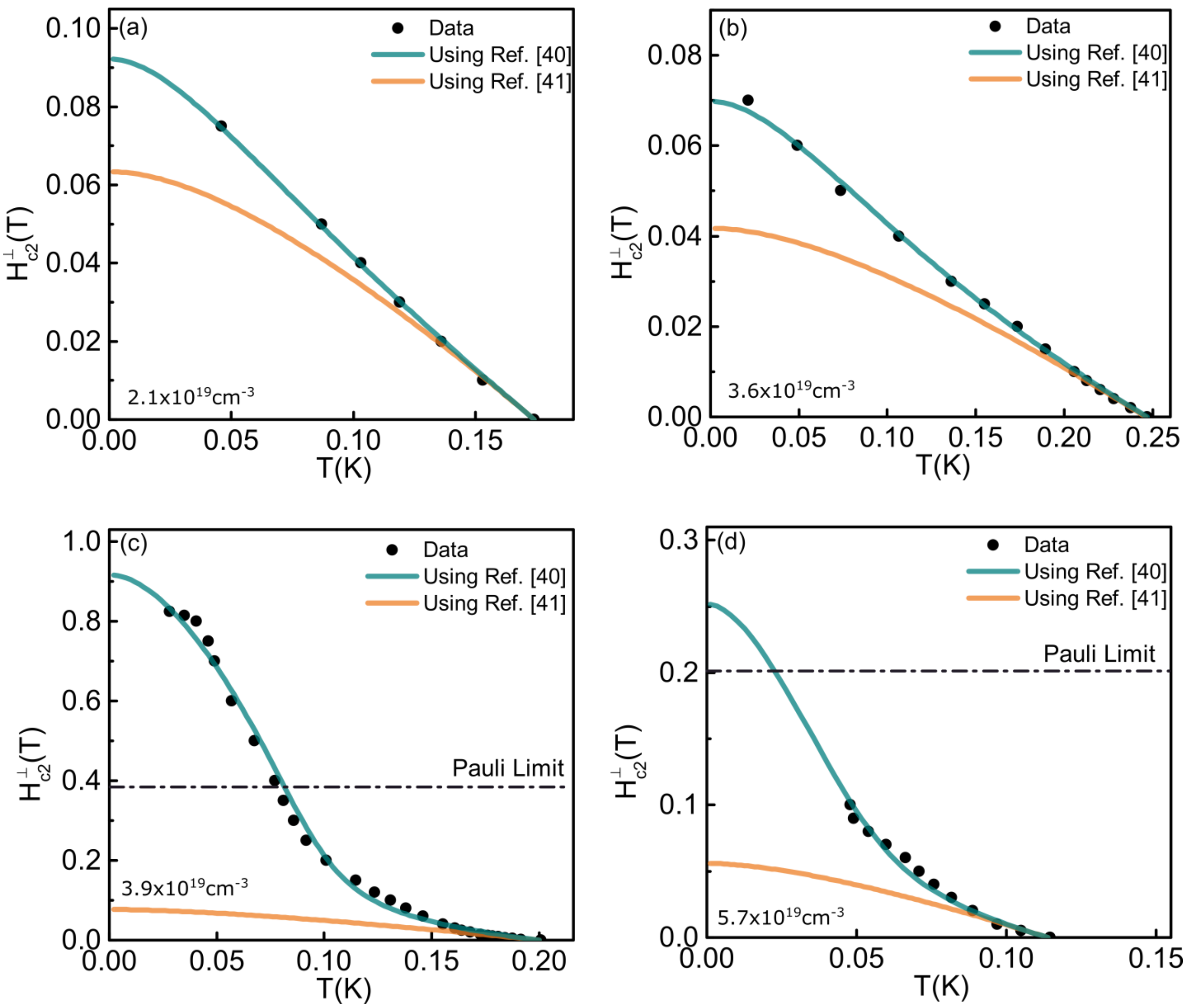

**Supplementary Fig. 2:** Out of plane critical field, $H_{c2}^{\perp}(T)$, for four samples, plotted as a function of temperature. The solid curves correspond to fits to our two-band model which includes orbital, Pauli and SO scattering effects, using estimated coupling constants from Ref. [42] ($\lambda_{11} = 0.3$, $\lambda_{22} = 0.1$ and $\lambda_{12} = \lambda_{21} = 0.015$) and from Ref. [41] (listed in the main text). The figures shows clearly that values of coupling constants from [42] do not produce a good fit, while those from [41] do. Our data are thus able to constrain the values of these parameters.

**Supplementary Table 1**: Table listing the best fit parameters to the data from the four samples, using the constraints from Refs. [41] and [42].

| | $n = 2.1\times10^{19}\,cm^{-3}$ (60-nm doped STO) | $n = 3.9\times10^{19}\,cm^{-3}$ (230-nm doped STO) | $n = 3.6\times10^{19}\,cm^{-3}$ (60-nm doped STO) | $n = 5.7\times10^{19}\,cm^{-3}$ (220-nm doped STO) |
|---|---|---|---|---|
| **Using coupling constants from Ref. [41]** | $D_1$=0.90 $cm^2/s$<br>$D_2$=3.3 $cm^2/s$<br>$\lambda_{so,1}$=2.0<br>$\lambda_{so,2}$=5.5<br>d=26 nm | $D_1$=0.036 $cm^2/s$<br>$D_2$=3.0 $cm^2/s$<br>$\lambda_{so,1}$=3.5<br>$\lambda_{so,2}$=0.4<br>d=8 nm | $D_1$=1.6 $cm^2/s$<br>$D_2$=7.4 $cm^2/s$<br>$\lambda_{so,1}$=2.5<br>$\lambda_{so,2}$=5.5<br>d=20 nm | $D_1$=0.16 $cm^2/s$<br>$D_2$=3.1 $cm^2/s$<br>$\lambda_{so,1}$=1.2<br>$\lambda_{so,2}$=0.6<br>d=17 nm |
| **Using coupling constants from Ref. [42]** | $D_1$=1.18 $cm^2/s$<br>$D_2$=2.1 $cm^2/s$<br>$\lambda_{so,1}$=2.0<br>$\lambda_{so,2}$=5.5<br>d=26 nm | $D_1$=0.19 $cm^2/s$<br>$D_2$=2.0 $cm^2/s$<br>$\lambda_{so,1}$=5<br>$\lambda_{so,2}$=4.8<br>d=11 nm | $D_1$=0.88 $cm^2/s$<br>$D_2$=4.52 $cm^2/s$<br>$\lambda_{so,1}$=2.5<br>$\lambda_{so,2}$=4.9<br>d=18 nm | $D_1$=0.124 $cm^2/s$<br>$D_2$=1.58 $cm^2/s$<br>$\lambda_{so,1}$=2<br>$\lambda_{so,2}$=1.56<br>d=17 nm |

## IV. Additional discussion of the R vs. T data

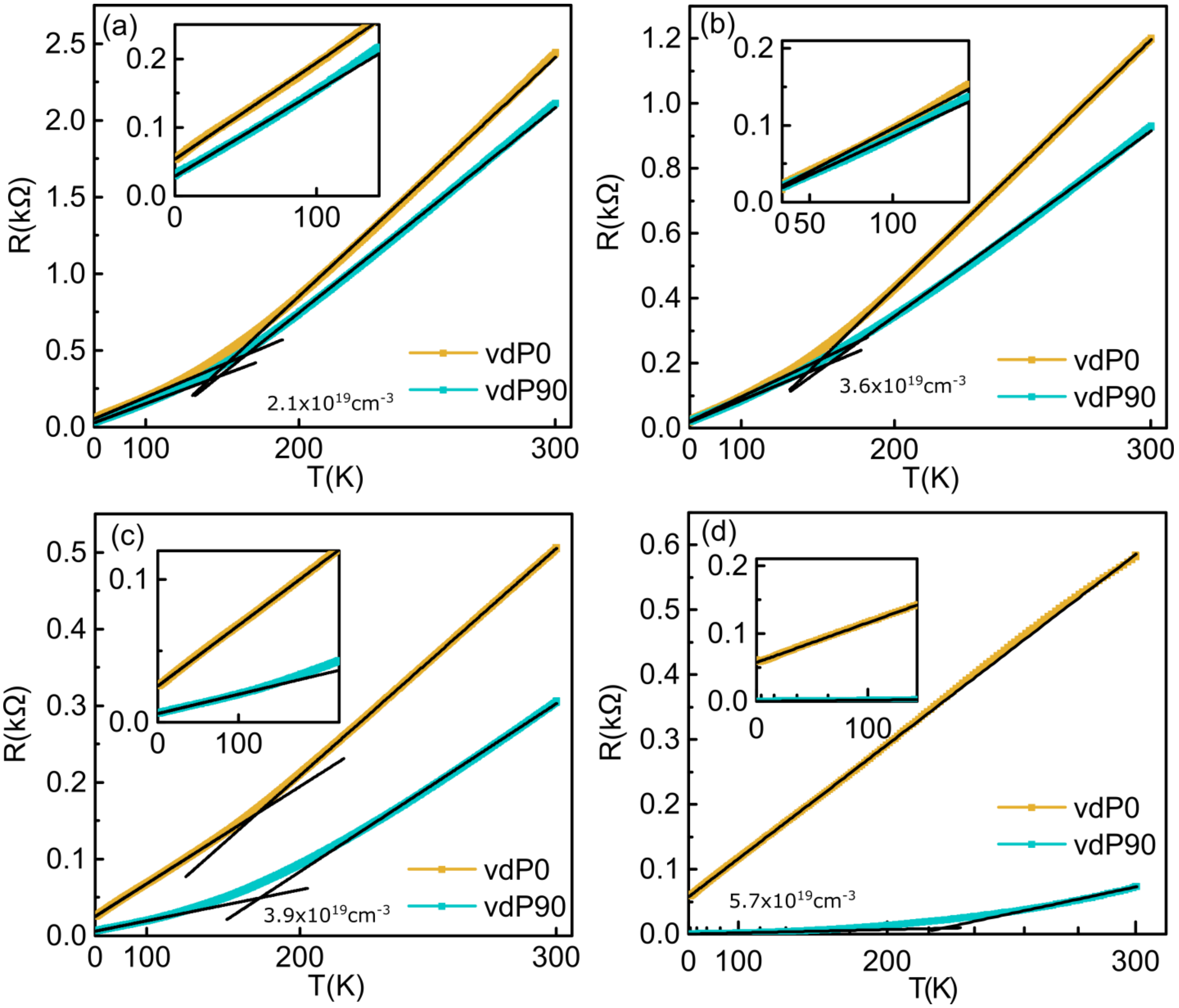

**Supplementary Fig. 3**: Normal state resistance as function of temperature (plotted on a quadratic scale) in the van der Pauw geometry for each sample in the main text. Shown are the results for the two possible ways to measure the resistance, differing by a 90-degree rotation of the current and voltage leads (see Supplementary Fig. 5). We refer to these two measurement configurations as vdP0 and vdP90, respectively. The solid lines correspond to fits using $R=R_0 + AT^2$, showing the lower T and higher T quadratic behaviors. Insets: the data zoomed in at lower T. The data shown in the main text is vdP0.

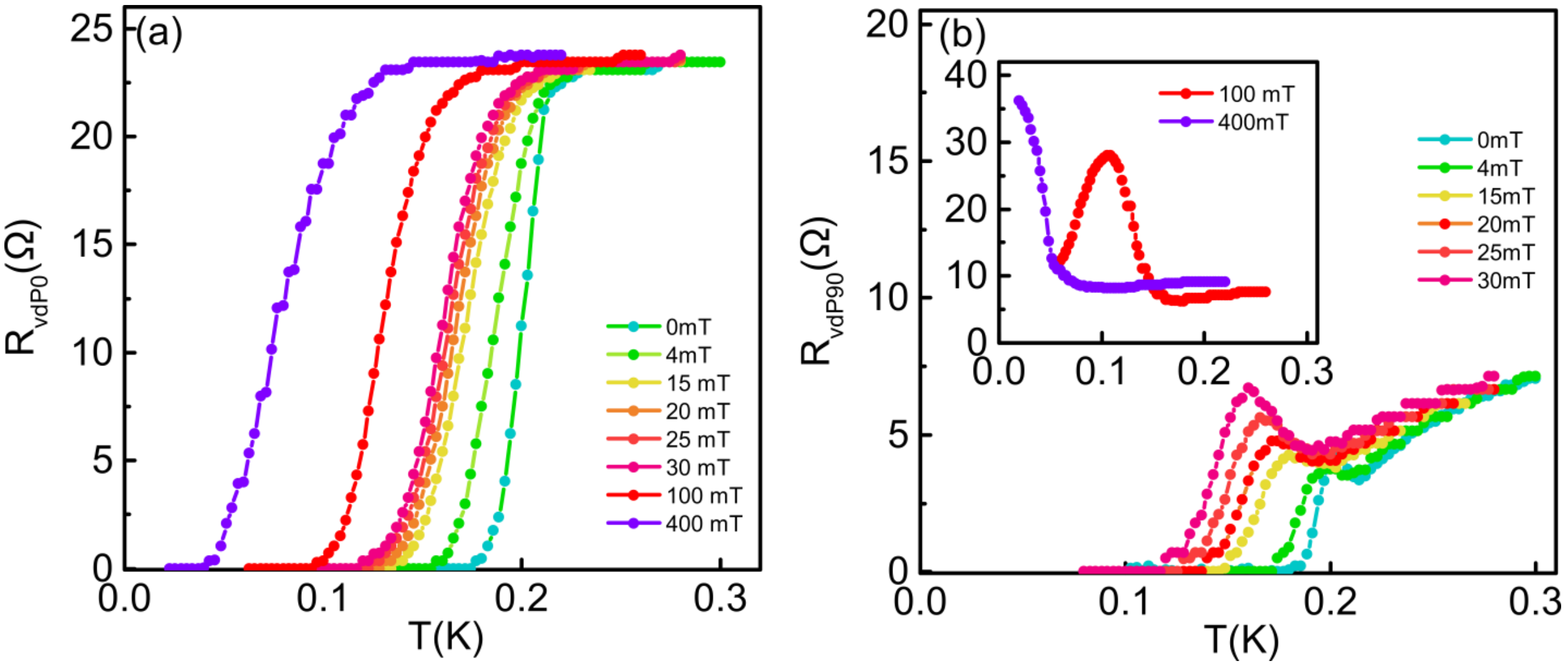

**Supplementary Fig. 4**: Resistance as function of temperature near the superconducting transition for different out-of-plane applied magnetic fields for the sample with $n = 3.9 \times 10^{19} cm^{-3}$ for the two possible van der Pauw configurations, (a) vdP0 and (b) vdP90. The data shown in the main text is vdP0. The peak near the superconducting transition measured as vdP90 is typical of all superconducting samples we measured and is analyzed in Section V, below. (The step-like behavior seen where the resistance varies slowly with temperature is due to discretization by the digital lock-in amplifier.)

**V. R vs. T curve analysis**

The upturn in R near the superconducting transition visible in Supplementary Fig. 4 at finite B is most likely due to spatial variations in resistivity across the sample, which are common for $SrTiO_3$-based heterostructures and other complex materials [S1–S5]. This type of transport feature has been observed in other superconductors and, excluding an exotic origin, can be analyzed within a simple resistor network model [S6], as illustrated schematically below (Supplementary Fig. 5).

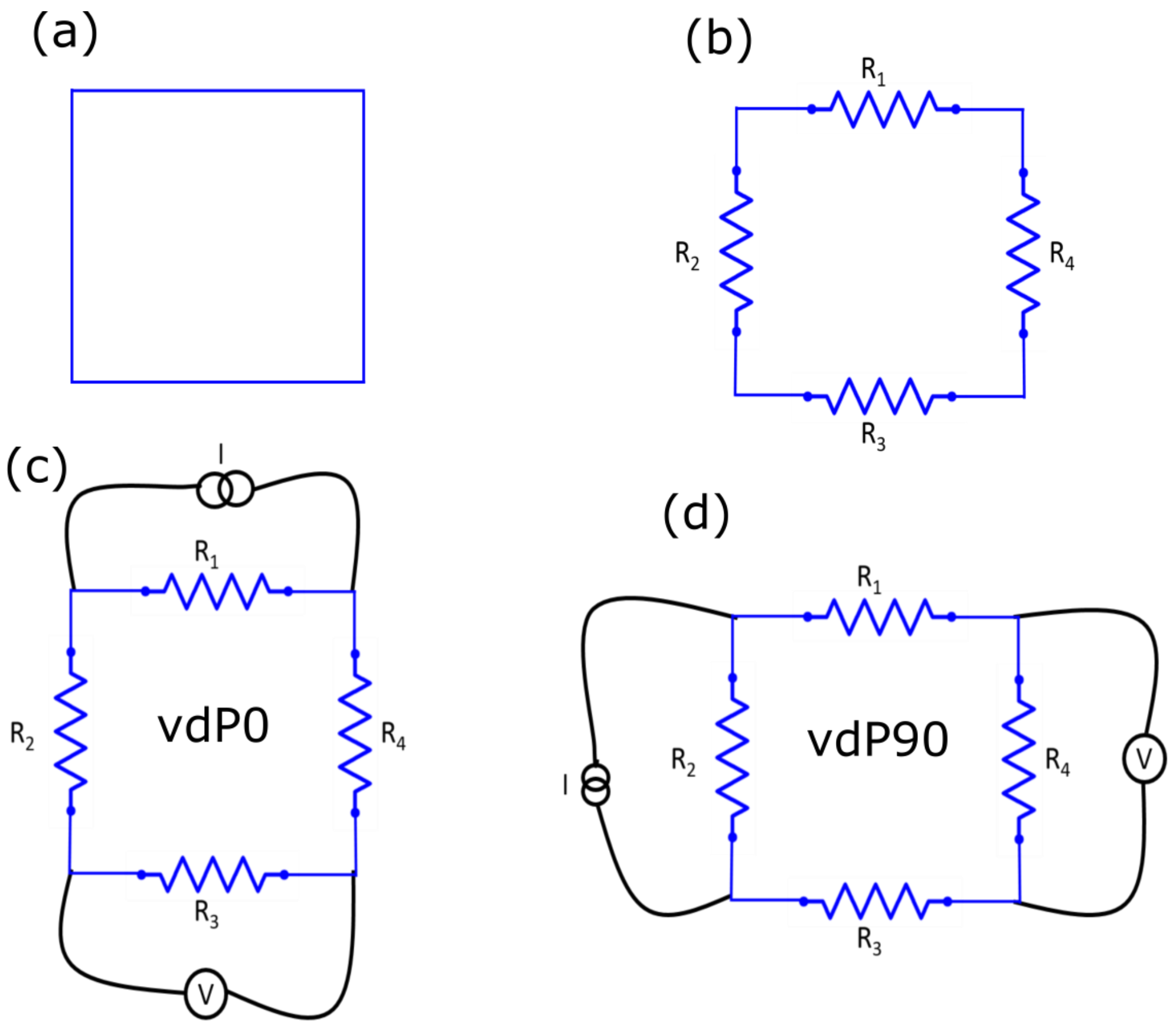

**Supplementary Fig. 5:** Schematics showing (a) a rectangular sample, (b) the sample being modeled as a four-resistor network, (c) one of the four-terminal van der Pauw measurement configurations (vdP0) and (d) the other configuration (vdP90), with current and voltage leads rotated by 90 degrees with respect to vdP0.

The voltage drops for a given applied current (I) for vdP0 and vdP90 respectively are:

$$V_{vdP0} = \frac{R_1 R_3}{(R_1+R_2+R_3+R_4)} I \qquad (1)$$

$$V_{vdP90} = \frac{R_2 R_4}{(R_1+R_2+R_3+R_4)} I \qquad (2)$$

The respective resistances are defined as $R_{vdP0} = V_{vdP0}/I$ , $R_{vdP90} = V_{vdP90}/I$ and $R_{Hall} = V_{Hall}/I = R_{vdP0} - R_{vdP90}$. If there is a difference in transition temperatures between $R_1$, $R_2$, $R_3$ and $R_4$, this leads to an upturn in the measured values of $R_{vdP0}$ (T) and $R_{vdP90}$(T) near the superconducting transition [S6]. For instance, if the critical temperatures of $R_2$ and $R_4$ are smaller than those of $R_1$ and $R_3$ an upturn is expected for the measured $R_{vdP90}$ from eqn. (2) above, while no upturn is expected for $R_{vdP0}$ according to eqn. (1). This is because $R_1$ and $R_3$ appear in the denominator of eq (2) but not on the numerator. This is consistent with our observation of an upturn along one measurement configuration and not the other.

To get a further qualitative and semi-quantitative insight, we make simplifying assumptions of $R_2$=$R_4$ and $R_1$=$R_3$, obtaining:

$$R_{vdP0} = \frac{R_1^2}{2(R_1+R_2)} \qquad (3)$$

$$R_{vdP90} = \frac{R_2^2}{2(R_1+R_2)} \qquad (4)$$

To illustrate this effect, in Supplementary Fig. 6 we calculate examples of $R_{vdP0}$(T) and $R_{vdP90}$(T) starting with generated $R_1$(T) and $R_2$(T) curves satisfying T(0.5*$R_{2\text{ normal state}}$) < T(0.5*$R_{1\text{ normal state}}$) for B>0, and using equations (3) and (4) to compute $R_{vdP0}$ and $R_{vdP90}$ from these curves. Note that the obtained curves match qualitatively very well with the experimental features from Supplementary Figs. 2 (a) and (b). We should add that experimentally, the upturn is always observed in the vdP configuration with the smaller normal-state resistance. In the context of the above resistor-model analysis, this is consistent with the fact that for a typical superconductor in the dirty limit the critical field scales with the normal state resistance ($H_c \propto R_{\text{normal state}}$) [3]. In other words, $T_c(R_2)$ ~ $T_c(R_1)$ (defined as half of the normal state resistance at B=0), that is T(0.5*$R_{2\text{ normal state}}$) ~ $T_c$(0.5*$R_{1\text{ normal state}}$) at B=0. Upon application of finite B, T(0.5*$R_{2\text{ normal state}}$) becomes less than T(0.5*$R_{1\text{ normal state}}$) because $H_c$ of $R_2$ is less than the $H_c$ of $R_1$ since $R_{2\text{ normal state}}$< $R_{1\text{ normal state}}$.

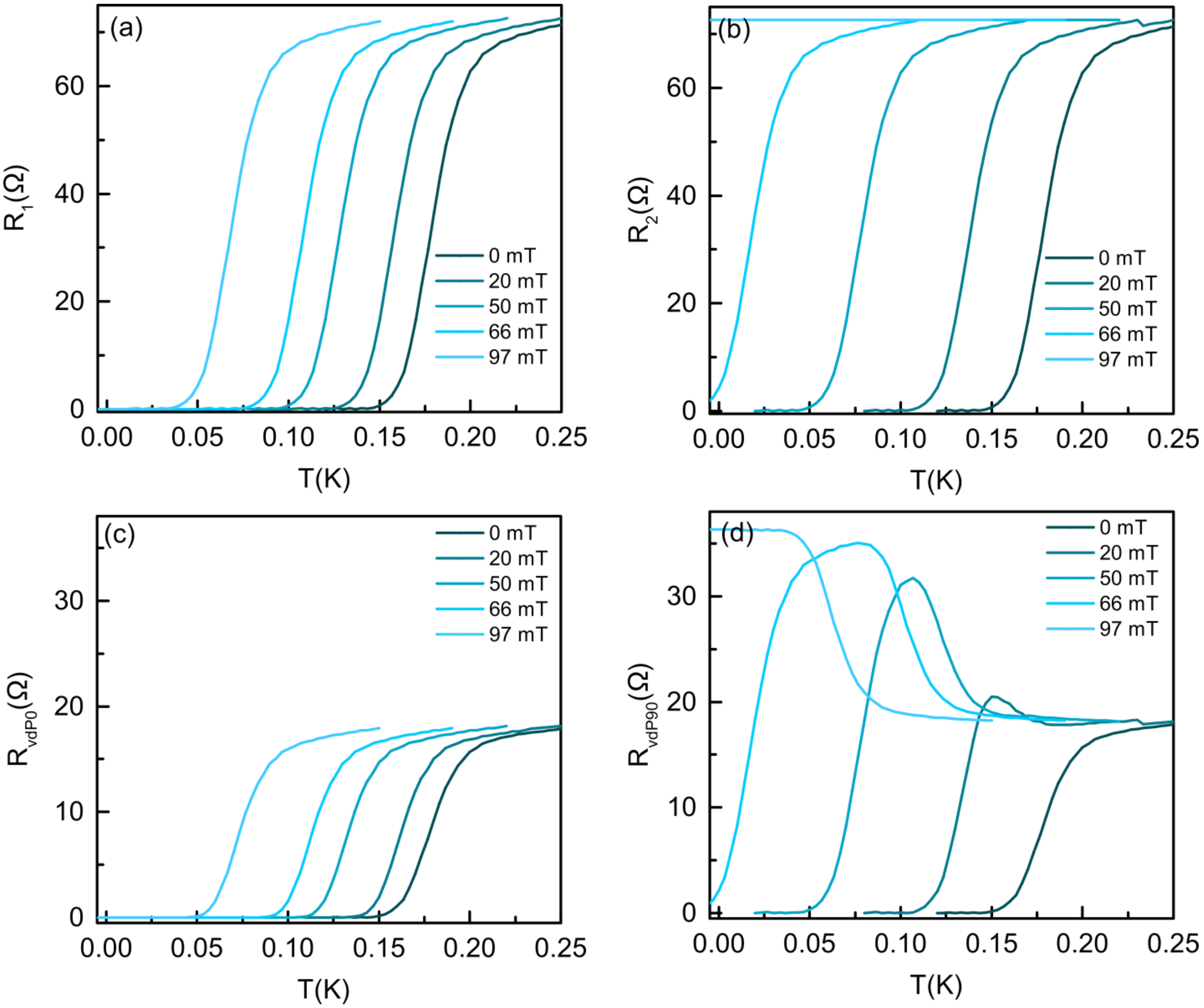

**Supplementary Fig. 6**: Generated R vs. T curves for (a) $R_1$=$R_3$ and (b) $R_2$=$R_4$ at different magnetic fields. The effect of an applied B field is included as a shift in the curves to lower $T_c$ as B increases, such that the $H_{c2}(T)$ for $R_1$ and $R_2$ follow a WHH curve. (c) $R_{vdP0}$ vs. T curves at different magnetic fields, computed from $R_1$ and $R_2$ from (a) and (b) using eqn. (3). (d) $R_{vdP90}$ vs. T curves at different magnetic fields, computed from $R_1$ and $R_3$ from (a) and (b) using eqn. (4).

## VI. Discussion of $H_c(T)$ analysis

Next we consider whether the observed differences between $R_{vdP0}(T)$ and $R_{vdP90}(T)$ have a bearing on the interpretation of $H_{c2}(T)$ discussed in the main text. In particular, we assess whether the non-negative $H_{c2}(T)$ curvature discussed in the main text and expected for a multi-band superconductor could arise adventitiously if the observed superconductivity were actually single-band and hence described by the WHH formalism. As shown in Supplementary Fig. 7, the answer to this question is negative: if the system were described by the single-band WHH model, which has negative $H_{c2}(T)$ curvature, then the measured $H_{c2}(T)$ for both vdP0 and vdP90 would also show negative curvature and be well-described by the WHH model, in contradiction to our observation. This demonstrates that the observed non-negative curvature is a real feature of the superconductivity in our samples. It also shows that different $H_{c2}(T)$ curves are expected along vdP0 and vdP90, each of which is indicative of the actual physical properties of the sample, but under different measurement conditions.

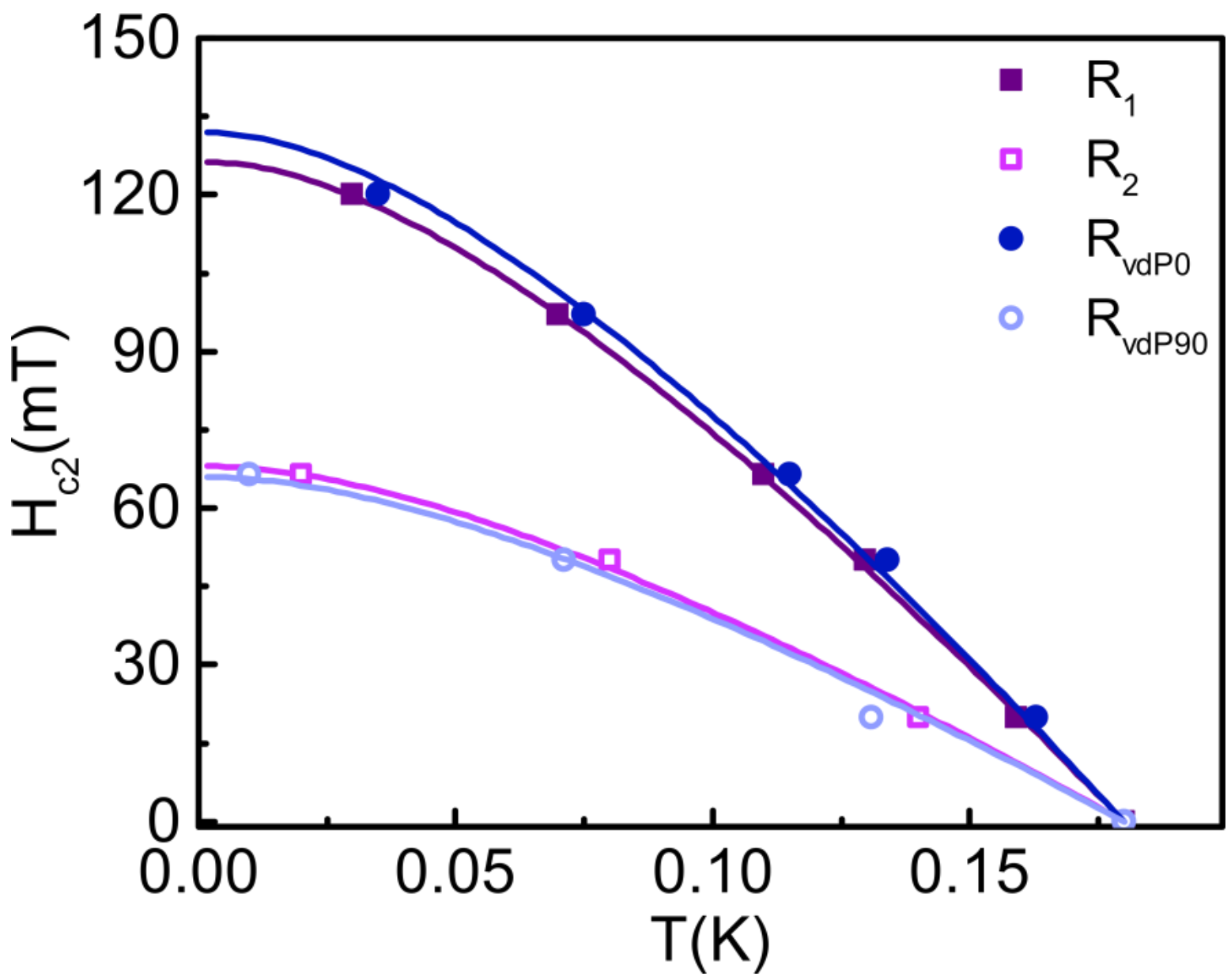

**Supplementary Fig. 7:** Critical field as a function of temperature corresponding to the synthetic $R_1(T,H)$, $R_2(T,H)$, $R_a(T,H)$ and $R_b(T,H)$ curves shown in Supplementary Fig. 6. For $R_1$ and $R_2$, the $H_{c2}(T)$ data points were picked to follow the WHH model, in order to mimic a single-band superconductor. In contrast, for $R_a$ and $R_b$, the $H_{c2}(T)$ data points were computed from the $R_a(T,H)$ and $R_b(T,H)$ curves, which were in turn obtained from $R_1(T,H)$ and $R_2(T,H)$ using eqns. (3) and (4). We note that these $H_{c2a}(T)$ and $H_{c2b}(T)$ data points also show a negative curvature, which fits well to the WHH model (solid curves). This indicates that the experimental non-negative curvature cannot be adventitious, and represents the underlying physics of the sample.

## VII. $H_{c2}$(T) for vdP90 for the sample with $n = 3.9\times10^{19} cm^{-3}$

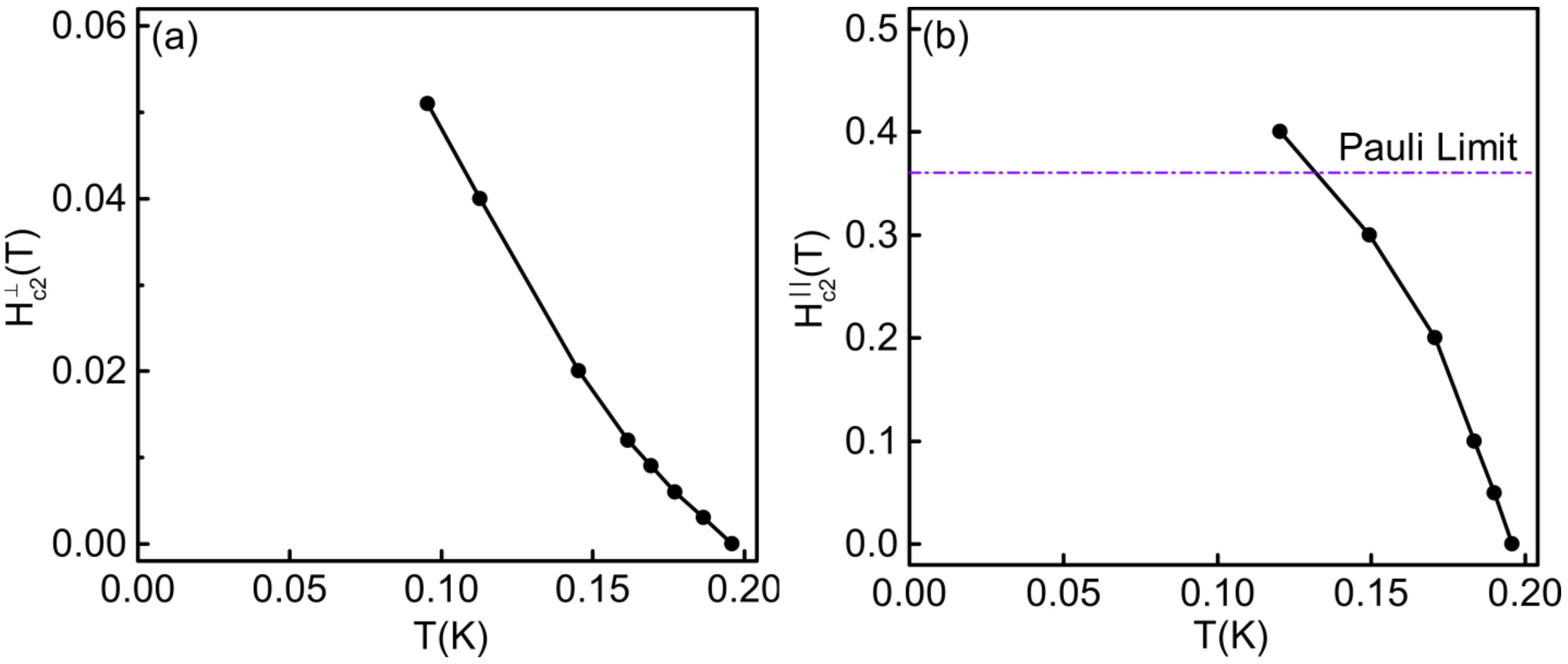

**Supplementary Fig. 8** Out-of-plane and in-plane critical fields as a function of temperature for the sample with $n = 3.9\times10^{19} cm^{-3}$, measured on vdP90 (data are from a separate cooldown from that in which the data in the main paper was taken). In this cooldown, it was not possible to reach temperatures below ~100 mK. (a) Out-of-plane critical field, $H^{\perp}_{c2}(T)$, showing non-negative curvature. (b) In-plane critical field, $H^{||}_{c2}(T)$, exceeding Pauli limit. The critical fields are determined by using the $0.5R_{\text{normal state}}$ criterion, where $R_{\text{normal state}}$ is the resistance before the onset of the peak. Note that the critical fields measured along vdP90 are about four times smaller than along vdP0 (Figs. 2(a) and 3(a) in the main text), consistent to the four times lower normal state resistance along vdP90 and the proportionality $H_c \propto R_{\text{normal state}}$ [3].

**Supplemental References:**